# A new route towards uniformly functionalized single-layer graphene

D W Boukhvalov[1,2] and M I Katsnelson[1]

[1]*Institute for Molecules and Materials, Radboud University Nijmegen*

*Heyendaalseweg 135, 6525AJ, Nijmegen, the Netherlands*

[2]*Computational Materials Science Center, National Institute for Materials Science, 1-2-1*

*Sengen, Tsukuba, Ibaraki 305-0047, Japan*

*It is shown, by DFT calculations, that the uniform functionalization of upper layer of graphite by hydrogen or fluorine does not change essentially its bonding energy with the underlying layers, whereas the functionalization by phenyl groups decreases the bonding energy by a factor of approximately ten. This means that the functionalized monolayer in the latter case can be easily separated by mild sonication. According to our computational results, such layers can be cleaned up to pure graphene, as well as functionalized further up to 25% coverage, without essential difficulties. The energy gap within the interval from 0.5 to 3 eV can be obtained by such one-side funtionalization using different chemical species.*

1. Introduction

Perspectives of broad industrial applications of graphene [1] in "electronics beyond silicon" and other fields [2] are crucially dependent on development of reasonably cheap and reliable methods of its mass production. Currently, this problem is still far from the solution. The first method of derivation of graphene, commonly used now to produce graphene sample for scientific research, is the mechanical exfoliation [1]. The sample size in this case can be as large [2] as 1 mm, with a very high sample quality. However, the output of monolayer graphene is very low which determines a very high cost of exfoliated graphene [3]. Another frequently used method is the production of epitaxial graphene by evaporation of surface layers of SiC with further deposition of graphene at the remaining surface [4-6]. The process requires rather restrictive temperature and time conditions; also, the sample quality, in particular, electron mobility is currently worse than for the exfoliated graphene. Recently, a prospective method has been proposed based on deposition of graphene on Ni(111) surface (the lattice mismatch in this case is very small) with further solution of nickel substrate [7]. Reduction of graphene oxide derived by solution of graphite in mineral acids with subsequent exfoliation is another promising way. A high degree of the reduction has been reached already [8,9], however, as discussed in the review [10] the reduction of graphene oxide to ideally clean graphene is rather problematic, due to a high energy of chemical binding of hydroxyl groups with graphene [11]. Also, graphene oxide can contain some perforations [12] or they can appear during the reduction process, similar to activation of the surface of graphite [13]. Other options are the solution of graphene without functionalization, in combination with sonication, with further exfoliation [14-16], or delamination of graphite intercalated by bromine [17]. However, transition from graphite to graphene in these cases is reversible and it is not easy to extract the monolayers from the solution containing also other graphene derivatives.

Thus, the main requirements to the methods of mass production of graphene are (1) high output of monolayers, (2) simplicity of their separation from other products of exfoliation, and (3) absence of oxygen, to avoid hardly reducible functionalization by hydroxyl groups [8,11].

For some (probably, the most promising) potential applications the functionalized graphene is better than the pristine one: pure graphene cannot be immediately used, for example, for transistors based on *p-n-p* junctions due to the Klein tunneling [18-20] and, thus, a controllable energy gap opening is desirable. Hydrogenation of graphene transforming it into graphane [21-23] is a prototype example of such functionalization. The energy gap in graphane turns out to be more than 5 eV [24], which is explained by transition of all carbon atoms from *sp2* to diamond-like *sp3* hybridization state. For the case of complete two-side coverage it is practically impossible to diminish the gap value just replacing of hydrogen by other chemical species [25]. On the contrary, one-side 25% functionalization of graphene (see Fig. 1a), as we will see below, can result in the energy gap of order of 1 to 2 eV, typical for conventional semiconductors. The one-side functionalization can be easily reached by putting graphene on SiO2 substrate [21]. The problem is that it is essentially inhomogeneous, the most probably, due to presence of ripples on graphene [26], as a result one has a strongly disordered semiconductor with a rather low electron mobility [21].

Thus, a high outcome of monolayer samples which can be easily cleaned from impurities is the main requirement to a method of mass production of graphene, and a stable *homogeneous* functionalization of the graphene monolayer is the main requirement to a method of production of graphene derivates with high electron mobility. Below we will discuss an approach which may be helpful in a solution of both these tasks. The basic idea is quite simple: if one functionalizes covalently the surface layer of graphite its binding energy with the underlying layer can be essentially diminished. The covalent functionalization of the surface of graphite by hydrogen and fluorine has been realized before [27], as well as the chemical functionalization of multi-wall carbon nanotubes (MWCNT) [28-31] and of epitaxial graphene on SiC substrate [32]. In the latter two cases, an insulator state has been reached. According to the band structure calculations [26,33] the metal-insulator transition in functionalized graphene takes place at relatively high degree of coverage, thus, observation of the insulating state for MWCNT and epitaxial graphene confirm an opportunity to reach a rather high coverage for graphene-like systems. Weakening of the bonds of the functionalized layer with bulk graphite makes probably possible its exfoliation by mild sonication, similar to what was realized for bromine-intercalated graphite [17]. A crucial difference with the exfoliation for intercalates is that for the functionalization of only one upper layer we avoid the problems with low outcome of monolayers and difficulties of separation from multilayers. Last not least, absence of the ripples in graphite (in contrast with graphene) means the absence of clusterization centres leading to inhomogeneous coverage [21,26].

## 2. Stability and energy gap of one-side uniformly functionalized graphene

First, we will estimate the value of energy gap at one-side functionalization with the coverage 25% and cohesion energy of the functional groups with graphene monolayer. High cohesive energy means that the functionalized graphene is suitable for use as a semiconductor, the low one means that it can be easily cleaned up to the pure graphene. At the second part of our work, we will estimate the binding energy of upper layer with the underlying one and formulate specific recommendations on the choice of chemical species for the functionalization.

The modeling is carried out by the density functional theory realized in the pseudopotential code SIESTA[31], as was done in our previous works[11,17,23,25,26,33]. Technical details of the calculations are presented in Ref. 25. All calculations are done in two versions, in the generalized gradient approximation (GGA) [35] and in the local density approximation (LDA)[36]. The former estimates more accurately the energy of covalent bonds[23] whereas the latter gives much better results for interlayer distances and interlayer binding energies in such systems[17,33]. An applicability of various functionals for description of layered systems is discussed in Ref. 33 (and refs. therein).

To model one-side 25% coverage of graphene we use the supercell with 8 carbon atoms where two of them are covalently bonded with different functional groups (see Fig. 1a,d and Fig. 2a), from hydrogen atom to complicated molecules similar to the groups used experimentally for functionalization of graphitic surfaces [29,37,38]. We consider here only adsorption of pairs of functional groups in the most energetically favourable positions. The adsorption of the pairs in other positions, as well as adsorption of single groups is not energetically favorable as was discussed in details in a series of previous works, see, e.g., Refs. 23, 25 and references therein. The functional groups in our simulations (alkenes) have been chosen in such a way that one can study dependence of the energy gap and chemisorption energy on the length of the group and presence of double bonds in it. One can see in Fig. 2b that the one-side uniform chemisorption on graphene with 25% coverage results in the opening of the energy gap. The larger the group, the smaller, in general, the energy gap is (although this dependence is not monotonous); for the groups with more than 5 atoms the gap value is less than 2 eV, that is, comparable with that in commonly used semiconductors. It is worthwhile to note that, although for the case of phenyl groups 25% coverage is less energetically favorable than 6.25% one, and the latter does not result in the opening of the energy gap, the energy difference is not so large, about 0.25 eV per group (which is close to the results of Ref. 39), that is, four times smaller than the activation energy which means that, probably, both configurations should be considered as stable. As usual, for the GGA calculations these data should be considered as estimations of the gap value from below; the more accurate GW calculations give larger values of the gap (see Ref. 24 and references therein). Changing the chemical composition, one can smoothly vary the energy gap value. It is worth to note that the stability decreases with the increase of number of atoms in the chemisorbed group and the presence of double bonds results in a decrease of the energy gap value.

To evaluate energetic stability of the compounds under consideration one should calculate the activation energy, that is, the energy of breakaway of one of the functional groups from functionalized graphene (see Fig. 1e), which is the initial stage of desorption [40]. The computational results are shown in Fig. 2c. The activation energy calculated by us for graphane is 4.2 eV. Keeping in mind that graphane can be completely transformed into pure graphene by annealing at 400°C [21], cleaning of the functionalized monolayers by moderate heating will be quite possible and not too difficult. At the same time, for all compounds under consideration the activation energy is much larger than 0.5 eV which means their stability at room temperature and moderate heating unavoidable for working electronic devices.

## 3. Weakening of binding in functionalized few layer graphene

At the last part of our paper we will consider possible ways of production of the functionalized graphene from graphite. We will simulate the functionalization of graphite by hydrogen and fluorine used earlier experimentally [27], as well as by methyl (-CH$_3$) and phenyl (-C$_6$H$_5$) groups. The latter may be considered as a simplified model of nitrophenyl (-C$_6$H$_4$NO$_2$) group used in Ref. 32 to functionalize the epitaxial graphene. In computations, we use supercell of graphene multylayer slab containing 8 carbon atoms per layer (Fig. 1a). The binding energy is calculated as $E_{bind} = E_{N-1}$(pure) + $E_1$(functionalized) - $E_N$(functionalized), where $E_N$ is the energy of $N$-layer graphene. To study the dependence of the binding energy on the number of underlying layers we use the slabs with 3, 6, and 10 graphene layers with the Bernal AB package like in the bulk graphite. The computational results presented in Fig. 3a show that for the case of 25% coverage by methyl or phenyl groups the binding energy is negative which means that such functionalization is energetically unfavorable and, thus, unstable. The negative binding energy means that bulk graphite with the functionalized surface layer is less favorable than separated functionalized single layer and pristine bulk graphite. Thus, the functionalization of graphite stops when a minimal positive value of the binging energy is reached. Experimental data on the functionalization of epitaxial graphene show coverage about 10% [32]. This maximal coverage can be reached only for single atom species, such as hydrogen or fluorine. For larger functional groups, the steric effect is important, related to the fact that each functional group requires a definite amount of space. It is worthwhile to note that, according to our calculations of optimized atomic positions, the corrugation of graphene sheets at the functionalization characteristic for the monolayer graphene takes place in graphite as well. Moreover, it involves also underlying layers, in an agreement with the experimental data [27]. It is the corrugation which makes energetically unfavorable 25% coverage by large enough groups such as $C_6H_5$, where the distortions of the underlying layers are 1.5 times stronger than for monoatomic species. Experimentally [29], the ratio of intensities of C1s XPS peaks corresponding to $sp^2$ and $sp^3$ hybridization state for epitaxial graphene functionalized by nitrophenyl groups also shows much smaller degrees of coverage than 25%. This illustrates a general statement that chemical properties of graphene and graphite can be essentially different [2]. It follows from the results presented in Fig. 3a that the interlayer binding energy for functionalized graphite is smaller than the corresponding value for graphite (35 meV per carbon atom [33]) and does not depend too much on the number of carbon layers in the slab. Thus, three-layer graphene can be used as a minimal model to study semi-quantitatively functionalization of bulk graphite. Instability of graphite with the surface maximally covered by multiatomic groups does not mean that the functionalization is impossible but just that the real degree of coverage will be always smaller than 25%. To study this opportunity we have carried out calculations for larger supercells, with 18 and 32 carbon atoms per graphene layer bonded with two functional groups (see Fig.

1 b,c), which means 11% and 6.25% coverage, respectively, for three-layer graphene as a model of graphite. One can see from the computational results shown in Fig. 3b that the binding energy of the functionalized layer with other two layers increases with the coverage decrease. For hydrogen and fluorine this effect is quite small but for methyl and phenyl it is very essential, due to decrease of corrugations in underlying layers. The functionalization by $CH_3$ groups becomes stable for the 11% coverage, and by $C_6H_5$ - for 6.25% coverage. In the latter case, the binding energy of the functionalized layer with the neighboring one is just 4 meV, almost order of magnitude smaller than in pure graphite and very close to the binding energy in bromine-intercalated graphite at its delamination [17]. The binding energy per 32-atom supercell is 0.128 eV which is much smaller than the activation energy for a given functional group. This means that, the most probably, exfoliated graphene monolayer at, e.g., mild sonication will remain functionalized at this degree of coverage. To open the energy gap, this graphene should be put on SiO2 substrate and after that the degree of the functionalization should be increased up to the maximum, that is, 25%.

## 4. Conclusions

Thus, functionalization of graphite by phenyl groups reduces the interlayer binding energy by a factor of approximately 10 which makes exfoliation of the upper layer by mild sonication relatively simple. Importantly, only single upper layer has this small binding energy, in contrast with the mild sonication of intercalated graphite. Presence of chemisorbed groups should prevent conglutination of the monolayers back to graphite. This allows us to hope that the outcome of monolayer graphene for this method will be high enough. To produce pure graphene, the functional groups can be afterwards eliminated by a moderate heating since the activation energy is not very high. The functionalization of graphite, in comparison with that of graphene, is supposed to be more homogeneous due to absence of the ripples working as centers of functionalization. One can hope therefore that the functionalized graphene will have both suitable value of the energy gap and high degree of order, thus, high enough electron mobility. It would be very interesting to try to check these predictions experimentally.

**Acknowledgements** The work is financially supported by Stichting voor Fundamenteel Onderzoek der Materie (FOM), the Netherlands.

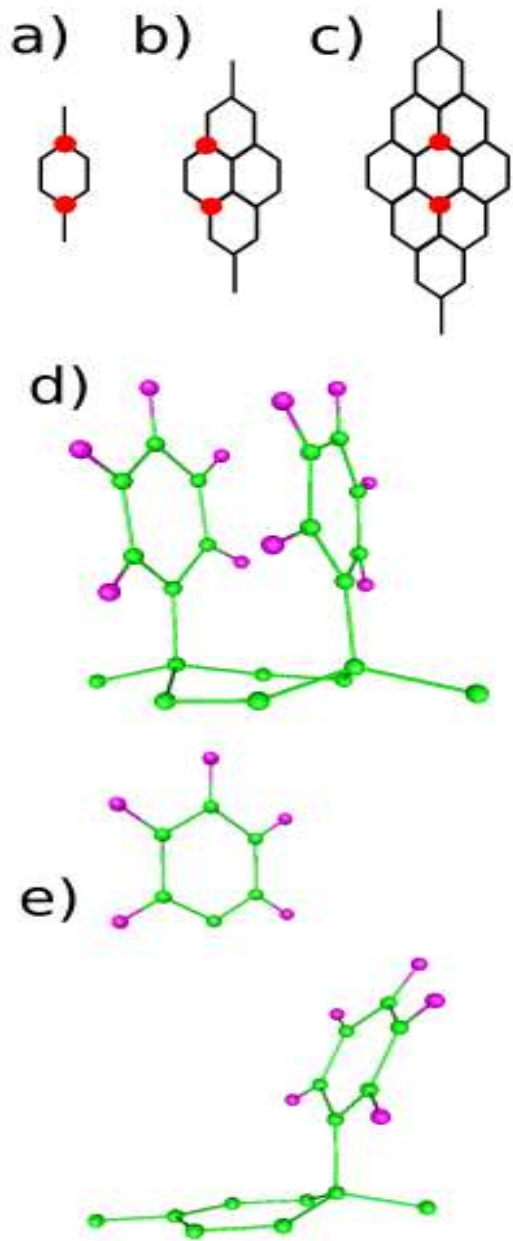

**Fig. 1** Supercells used for modeling of graphene with chemisorbed functional groups, for 25% (a), 11% (a), and 6.25% (c) coverage; (d) optimized geometry of the supercell of graphene with 25% coverage by phenyl groups; (e) optimized geometry of the activation of the latter system (disconnection of one of the phenil groups).

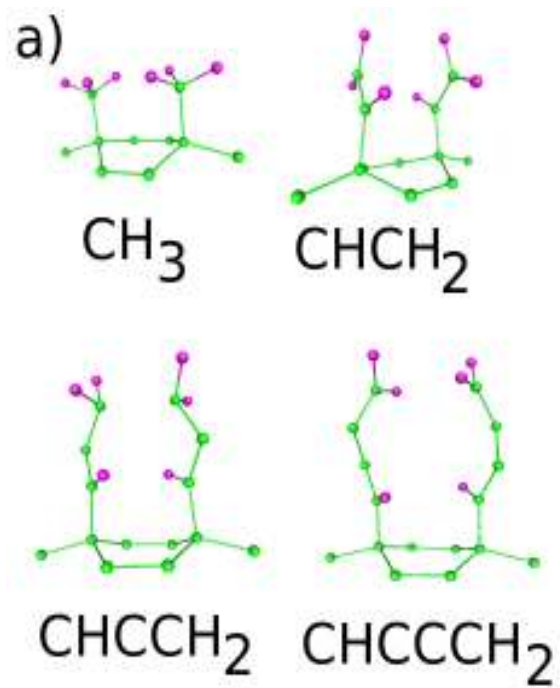
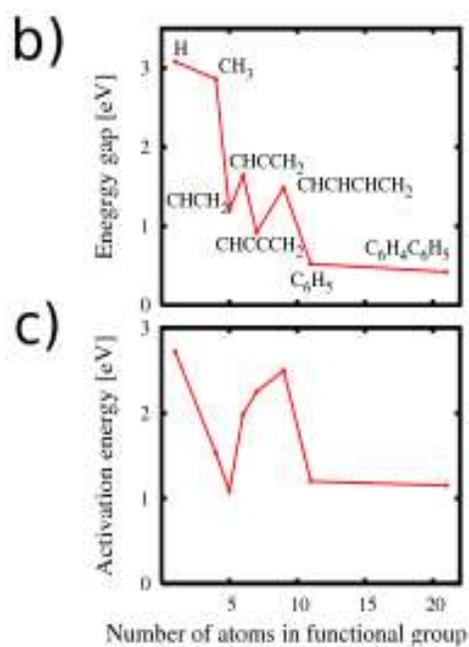

**Fig. 2.** Optimized geometry of the supercells of functionalized graphene (a); dependence of the energy gap (b) and activation energy (c) on functional groups chemisorbed on graphene with 25% one-side coverage.

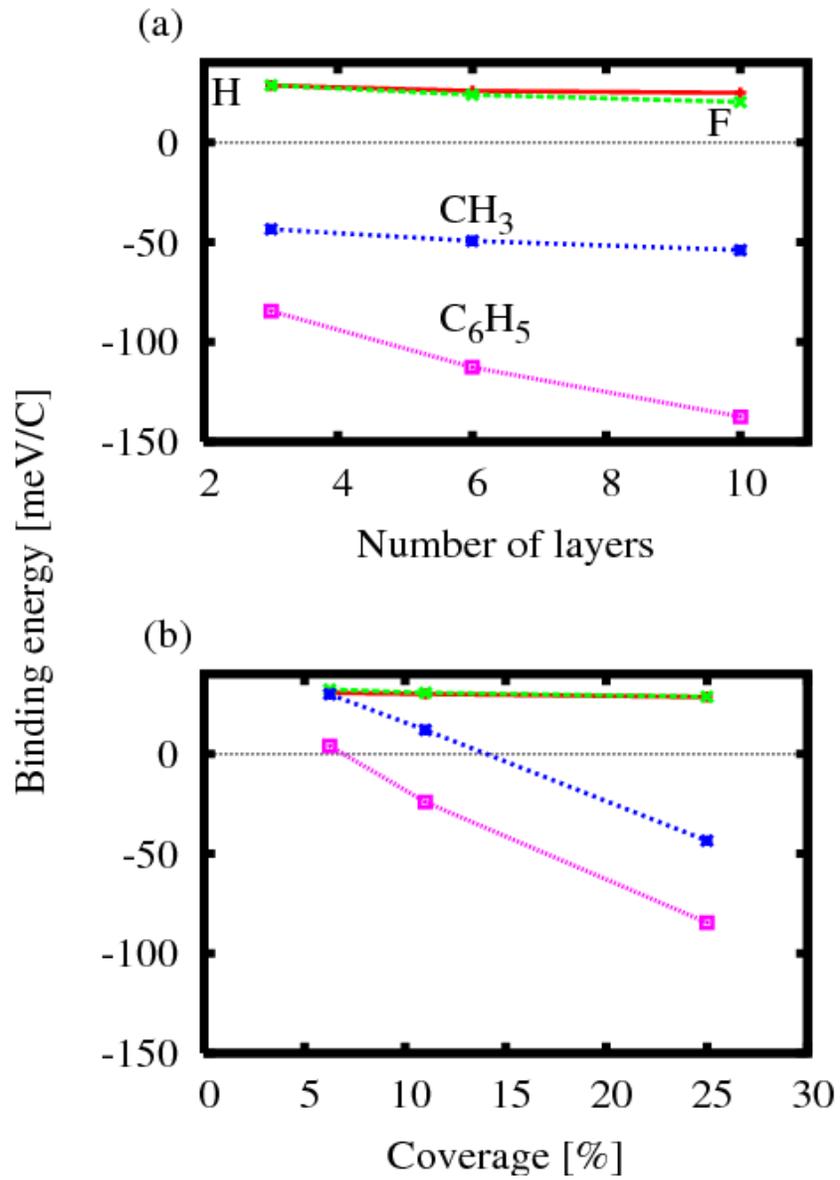

**Fig. 3.** Binding energy per carbon atom in graphene sheet between graphene layers as function of number of layers for 25% coverage (a) and coverage degree for three layer graphene (b).